\documentclass[prl,aps,twocolumn,longbibliography]{revtex4-2}
\usepackage{color}
\usepackage{graphicx}
\usepackage{bm}
\usepackage{amsmath}
\usepackage{amssymb}
\usepackage{amsthm}
\usepackage{amsfonts}
\usepackage{enumerate}
\usepackage{latexsym}
\usepackage{lipsum}
\usepackage[colorlinks=true,urlcolor=blue,citecolor=blue,allcolors=blue]{hyperref}

\usepackage{braket}
\usepackage{array}
\usepackage{float}
\usepackage{cleveref}

\DeclareMathOperator{\tr}{tr}
\DeclareMathOperator{\re}{Re}
\DeclareMathOperator{\im}{Im}

\begin{document}

\title{Theory of the Loschmidt echo and dynamical quantum phase transitions in disordered Fermi systems}
\author{Tuomas I. Vanhala and  Teemu Ojanen}
\affiliation{Computational Physics Laboratory, Physics Unit, Faculty of Engineering and
Natural Sciences, Tampere University, P.O. Box 692, FI-33014 Tampere, Finland}
\affiliation{Helsinki Institute of Physics P.O. Box 64, FI-00014, Finland}

\begin{abstract}
In this work we develop the theory of the Loschmidt echo and dynamical
phase transitions in non-interacting strongly disordered Fermi systems
after a quench. In finite systems the Loschmidt echo displays zeros in
the complex time plane that depend on the random potential
realization. Remarkably, the zeros coalesce to form a 2D manifold in
the thermodynamic limit, atypical for 1D systems, crossing the real
axis at a sharply-defined critical time. We show that this dynamical
phase transition can be understood as a transition in the distribution
function of the smallest eigenvalue of the Loschmidt matrix, and
develop a finite-size scaling theory. Contrary to expectations, the
notion of dynamical phase transitions in disordered systems becomes
decoupled from the equilibrium Anderson localization transition. Our
results highlight the striking qualitative differences of quench
dynamics in disordered and non-disordered many-fermion systems.
\end{abstract}

\maketitle

\emph{Introduction---} Equilibrium statistical physics is one of the
most general theories in natural sciences - it has been successfully
applied to a remarkably wide variety of systems between the smallest
and the largest scales in the universe.
Only recently, the advent of modern quantum simulators and digital
quantum computers has enabled a detailed experimental access to
coherent far-from equilibrium quantum evolution
\cite{Tian2020,Xu2020,Eckardt_2017,Flaeschner2018,Fogarty_2020,Guo2019,Tian2019,Wang2019,Flaeschner2018,Zhang2017,Jurcevic2017}.
One aspect of this that has recently stimulated enormous interest is
the possibility for non-analytic behaviour generated by a sudden
quench. This phenomenon, often discussed in terms of a many-body
Loschmidt echo, has close analogies with equilibrium phase transitions
which give rise to well-known non-analytic properties as a function of
the control parameter driving the transition. In contrast, a so-called
dynamical quantum phase transition, taking place at a critical time
$t_c$, signifies a vanishing Loschmidt echo and an abrupt change in
the temporal evolution \cite{Heyl_2018}. The possibility of
non-analytic evolution is in itself intriguing, however, the formal
analogy with equilibrium criticality has launched a search for
possible universal properties in far-from-equilibrium systems
\cite{Heyl_2018, Heyl2015}.

In this work we establish the theory of Loschmidt echo and dynamical
quantum phase transitions in non-interacting disordered many-fermion
systems. We discover that the singular dynamics of disordered Fermi
systems constitute a radical departure from the previously studied
many-body quenches. We find that
i) the temporal evolution of the studied system after a generic quench
is accompanied by a vanishing Loschmidt echo after a finite time
ii) the critical time, when the Loschmidt echo vanishes, becomes a
deterministic non-fluctuating quantity in the thermodynamic limit
iii) the Loschmidt echo remains strictly zero after the critical time
iv) the qualitative behaviour of the Loschmidt echo does not depend on
whether the quench crosses the equilibrium Anderson localization
critical point or not.
Many insights on dynamical phase transitions have been obtained from
free-fermion systems and subsequently confirmed in a number of
strongly-correlated systems. Thus, our present work provides a
baseline to understand singular dynamics of even more complex
disordered systems in the future.

\begin{figure}
\includegraphics[width=1.0\columnwidth]{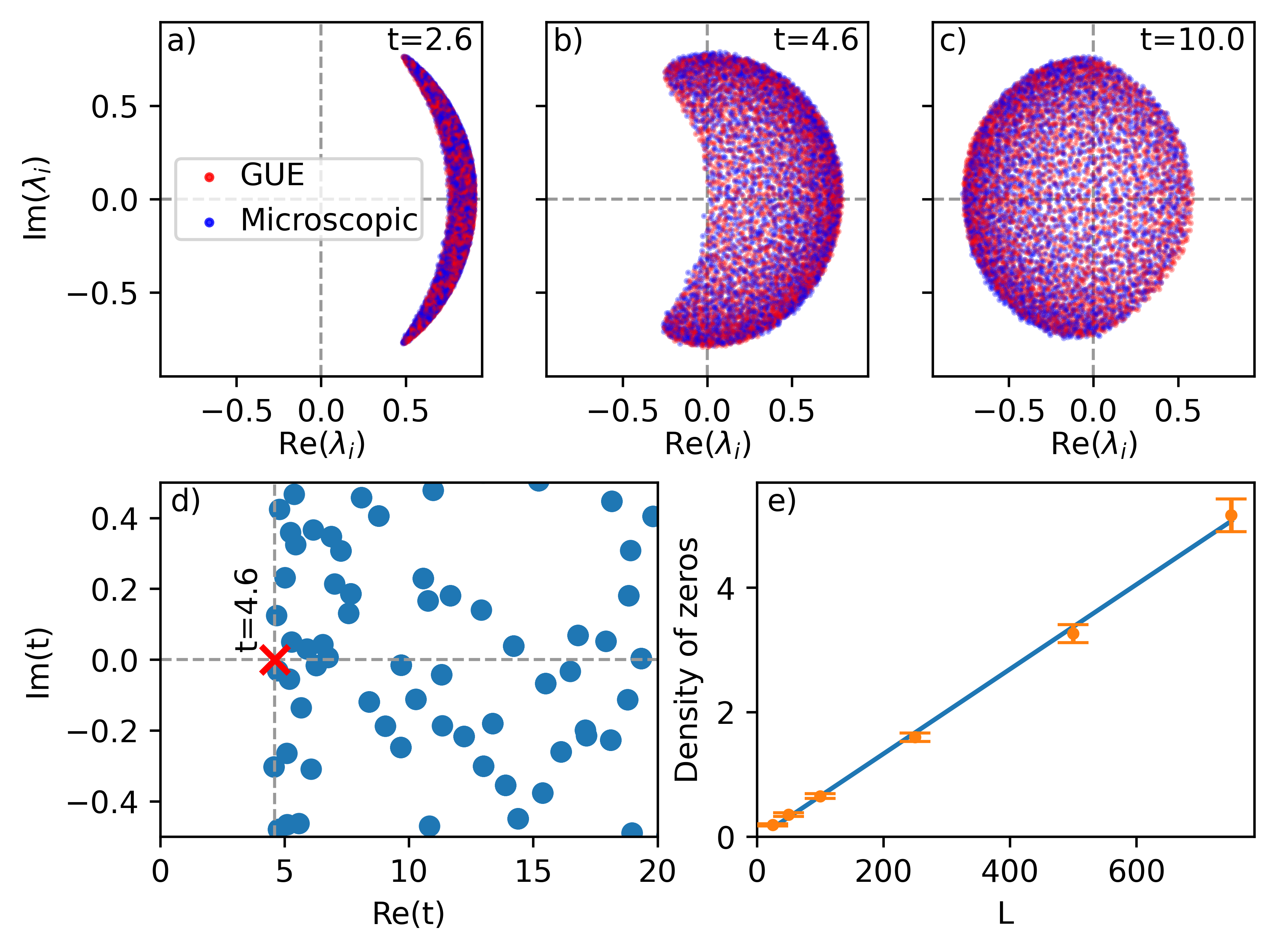}
\caption{Dynamical phase transition following a very strong sudden
  quench.
a)-c) Eigenvalues of the $M(t)$-matrix for two different models, the
GUE model and the microscopic model, which exhibit remarkable
similarity. The dynamical phase transition occurs at the point $t=t_c$
where the boundary of the eigenvalue distribution crosses the origin.
d) Loschmidt zeros for a single realization of the microscopic
model. The red cross marks the dynamical phase transition point where
the real axis intersects the boundary of the area with a finite
zero-density.
e) Scaling of the average density of zeros with system size $L$ in the
region $15<\re(t)<25$, $|\im(t)|<0.2$. The datapoints were calculated
with $100$ realizations of the random potential for each $L$ and the
line is a linear regression.}
\label{fig1}
\end{figure}

\emph{Prototype model and quench protocol---} We consider sudden
quenches between two generic, non-interacting fermionic Hamiltonians
$H_0$ and $H_1$. The initial state $\ket{\psi}$ of the system at time
$t=0$ is taken to be an $N_p$-particle eigenstate of $H_0$, which is
then propagated by $H_1$. Since both $H_0$ and $H_1$ are
non-interacting, the state is a Slater determinant for all times.
Collecting the $N_p$ occupied orbitals in the initial state as columns
of the matrix $V$, the evolved state $\ket{\psi(t)}$ is represented by
a Slater determinant of the columns of $V(t)=\exp(-i t H_1) V$, while
the associated Loschmidt echo is given by \cite{plasser2016efficient}
\begin{equation}
  Z(t) = \braket{ \psi | \psi(t) } = \det( M(t) ),
  \label{Z_M_first_definition}
\end{equation}
where the Loschmidt matrix $M(t)$ is defined as $M(t)=V^\dagger
V(t)$. It is convenient to consider the echo in the eigenbasis of
$H_1$ represented as $H_1=U_1 E_1 U_1^\dagger$, where columns of $U$
are the eigenstates of $H_1$ and $E_1$ is a diagonal matrix of the
eigenenergies. The matrix $M(t)$ can be written as
\begin{equation}
M(t) = V^\dagger U_1 \exp(-i t E_1) U_1^\dagger V,
\end{equation}
and the echo is determined by the basis change matrix $V^\dagger U_1$
and the distribution of the energies $E_1$.

As a prototype we consider a 1D Anderson model with a second quantized
Hamiltonian of the form
\begin{equation}
  H=J \sum_{i=1}^L (c_i^\dagger c_{i+1} + c_{i+1}^\dagger c_i) + \sum_{i=1}^L h_i c_i^\dagger c_i.
  \label{microscopic_model_hamiltonian}
\end{equation}
It is instructive to first consider the extreme quench where the
initial Hamiltonian $H_0$ is free from disorder $h_i=0, J=1$ while the
final Hamiltonian $H_1$ has no hopping $J=0$ and $h_i$ are drawn from
a uniform distribution in the interval $[-h/2,h/2]$. We quench from an
eigenstate of $H_0$, randomly choosing $N_p$ single-particle states to
be occupied, and propagate with the Hamiltonian $H_1$, whose spectrum
is just given by the $h_i$. We always consider half-filling,
$N_p=L/2$.

It is revealing to compare the microscopic model with an effective
random matrix model where the initial hamiltonian $H_0=U_0 E_0
U_0^\dagger$ is drawn from the gaussian unitary ensemble (GUE). In
this model it makes no difference what the eigenstates of $H_1$
are. This is because $U_0$ is distributed according to the circular
unitary ensemble (CUE) and, by the invariance of the Haar measure, so
is the overlap matrix $U_0^\dagger U_1$ regardless of what $U_1$
is. The matrix $V^\dagger U_1$ is thus just a randomly chosen
collection of row vectors from a CUE random matrix regardless of
$U_1$. We still assume that the energies $E_1$ are the same as in the
microscopic model, i.e. uniformly distributed in the interval
$[-h/2,h/2]$.

Dynamical phase transitions are revealed by the zeros of $Z(t)$,
termed the Loschmidt zeros, in the complex $t$-plane. A dynamical
phase transition takes place at a critical time $t_c$, where the
manifold of zeros intersects the real time axis. However, first we
explore the Loschmidt echo by following the evolution of the
eigenvalues of $M(t)$ and plotting them in the upper row of
Fig. \ref{fig1}. As $Z(t)$ is the determinant of $M(t)$, zeros of
$Z(t)$ appear at points $t$ where $M(t)$ has eigenvalue zero. We
notice that the spectrum of $M(t)$ falls within a well-defined,
bounded region at all times. After the appearance of the first zero
eigenvalue, the spectrum of $M(t)$ encapsulates the origin at all
times. This signifies a type of singular many-body dynamics where the
Loschmidt echo vanishes for all times after first reaching zero and
the only singularity of the rate function is at the critical time,
similarly to the quench within the massless phase in the 2D Kitaev
model \cite{PhysRevB.92.075114}. Here we observe a certain
universality, as the microscopic model and the GUE model produce
essentially the same eigenvalue distribution.

We can now confirm the nature of the dynamical phase transition also
by directly looking at the Loschmidt zeros which we locate using the
cumulant method developed in \cite{PhysRevX.11.041018}. We discuss the
specialization of the method to the non-interacting case in the
supplementary material \cite{supplemental_material}. In this case, as
also clarified below, the Loschmidt zeros are organized as a 2D
manifold. As seen in the lower left panel of Fig. \ref{fig1}, the
zeros indeed cross the real time axis at the time $t_c$ which
coincides with the first appearance of zero eigenvalue of $M(t)$. The
density of the Loschmidt zeros in the region $t>t_c$ increases
proportionally to $L$, which we also verify numerically in
Fig.~\ref{fig1}e. The zeros thus form a two-dimensional manifold
previously found in two-dimensional models \cite{PhysRevB.92.075114},
while one-dimensional models typically display lines of zeros
\cite{Heyl_2018}. Here, however, the zero manifold is two-dimensional
for both the 1D microscopic model and for the random matrix model
where the propagating Hamiltonian only enters through its
eigenenergies, which are not directly related to its dimensionality.

\begin{figure}
\includegraphics[width=1.0\columnwidth]{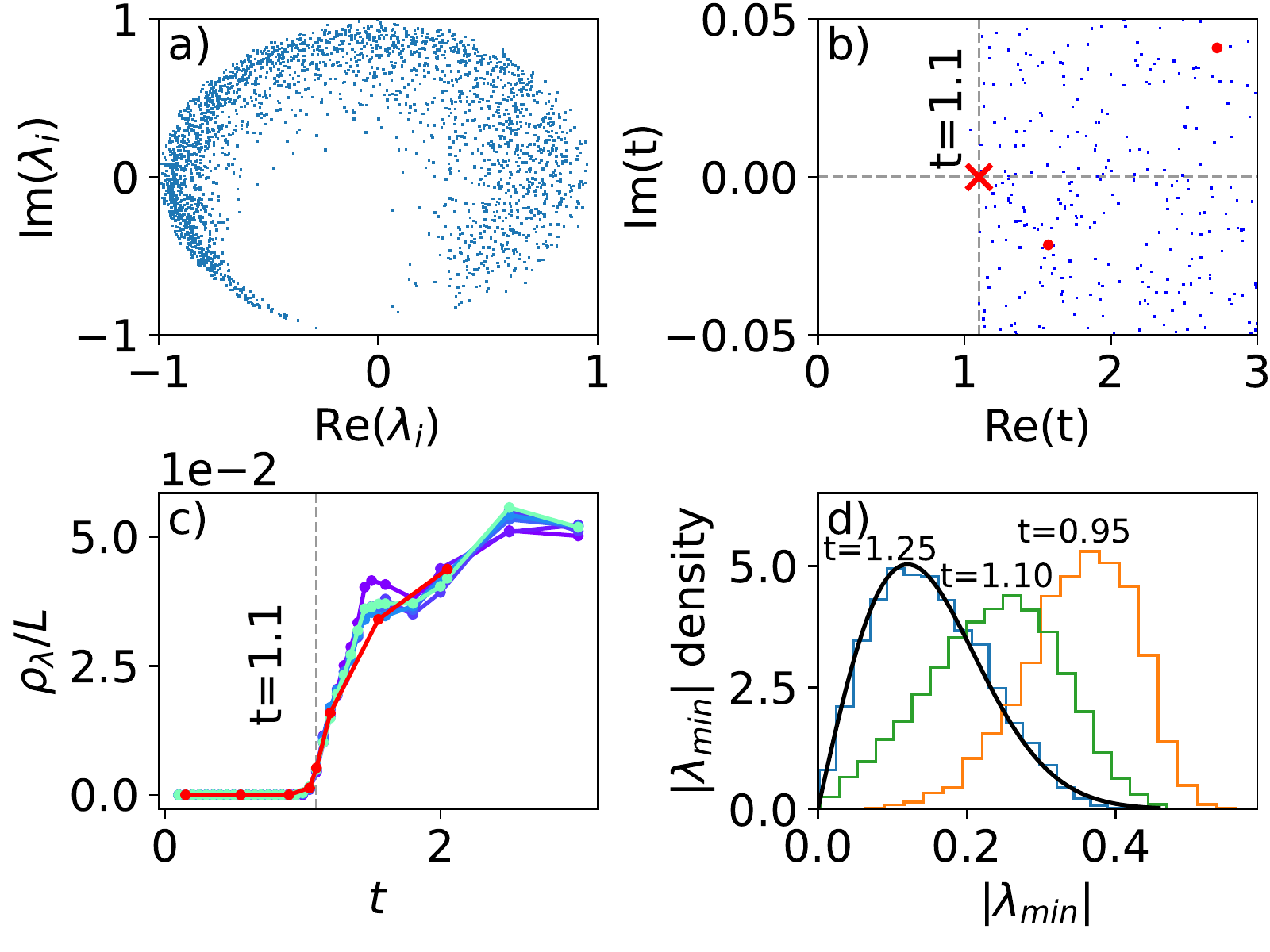}
\caption{Quenching from the ground state of the $(h=0,J=1)$ system to
  the $(h=5,J=1)$ system.  a) Eigenvalues of $M(t)$ for a single
  potential realization in a system of size $L=5000$ at the
  approximate critical time $t=1.1$.  b) Combined Loschmidt zeros for
  system size $L=500$ and $N_s=100$ potential realizations. The red
  dots highlight the zeros for one realization, showing their
  rarity. The red cross indicates the dynamical phase transition. c)
  The density of eigenvalues of $M(t)$ within a small disc of radius
  $0.05$ around the origin averaged over a large number of potential
  realizations. The color coding for system sizes and the sample sizes
  are as in Fig.~\ref{mean_std_figure}.  d) Histogram of $N_s=10000$
  samples of $|\lambda_{min}|$ for different times in the $L=500$
  system. The black line is a Rayleigh distribution fitted to the data
  at $t=1.25$.}
\label{h_0_to_5_quench_fig}
\end{figure}

\emph{Eigenvalue phase transition and scaling theory---} In the case
of extreme quenches between clean and non-hopping states discussed
above, the eigenvalue distribution of $M(t)$ appears to have a sharp
boundary and the critical time $t_c$ is easily located by simply
following the evolution of eigenvalues in the complex plane. However,
for generic quenches this is not the case, as the boundary of the
eigenvalue distribution in the thermodynamic limit may be difficult to
determine from the finite set of eigenvalues calculated for some
attainable system size $L$. Indeed, the question remains whether such
a sharp boundary generally exists even in the thermodynamic
limit. This is demonstrated in Fig.~\ref{h_0_to_5_quench_fig}a for a
quench starting from the ground state of Hamiltonian
\ref{microscopic_model_hamiltonian} at parameters $(h=0,J=1)$ and
propagated with $(h=5,J=1)$.

To gain a quantitative understanding of the dynamical phase
transitions in such cases, we consider ensembles of quenches for
different realizations of the random potential. In
Fig.~\ref{h_0_to_5_quench_fig}b we plot the combined Loschmidt zeros
for $100$ systems of length $L=500$. Although the zeros for a single
system are very sparse, the ensemble reveals a sharp boundary where
the zeros appear. From the viewpoint of the eigenvalues of $M(t)$, the
same boundary is found by inspecting the density $\rho_{\lambda}$ of
eigenvalues close to the origin averaged over the ensemble (see
Fig.~\ref{h_0_to_5_quench_fig}c). But is this sharp transition a
property of the ensemble, or a property of an individual quench in the
thermodynamic limit? If we could increase the system size
sufficiently, would we obtain a sharp non-fluctuating critical time
for an individual quench? To answer this question, we propose to study
the distribution of $|\lambda_{min}|$, the absolute value of the
smallest eigenvalue of $M(t)$. Assuming $M(t)$ is diagonalizable, this
is the same as the smallest singular value of $M(t)$, which has also
been studied in classical random matrix ensembles
\cite{Tao2010,BENAYCHGEORGES2012120}.

Suppose now that a sharp boundary for the eigenvalue distribution of
$M(t)$ exists in the thermodynamic limit. Then, for $t<t_c$, the
distribution of $|\lambda_{min}|$ should become increasingly narrow
with increasing $L$, and its mean $\left \langle |\lambda_{min}|
\right \rangle$ should approach a finite value corresponding to the
distance from the origin to the boundary of the eigenvalue
distribution.
For times $t>t_c$, on the other hand, both the mean and the standard
deviation $\sigma(|\lambda_{min}|)$ of the distribution should scale
to zero with increasing $L$. If we assume that the eigenvalues of
$M(t)$ in the vicinity of the origin are drawn independently from some
smooth distribution with local density $\rho_{\lambda}$, we find that
the distribution of $|\lambda_{min}|$ converges to the Rayleigh form
\cite{supplemental_material}
\begin{equation}
f(|\lambda_{min}|)=2 \pi r \rho_{\lambda} \exp(-\pi \rho_{\lambda} |\lambda_{min}|^2).
\end{equation}
As it is expected that $\rho_{\lambda}$ is proportional to $L$, the
mean and standard deviation can both be calculated to be proportional
to $1/\sqrt{L}$. We therefore expect a phase transition in the
distribution of $|\lambda_{min}|$ at $t_c$ where the scaling behaviour
of $\left \langle |\lambda_{min}| \right \rangle$ as a function of $L$
changes.

\begin{figure}
\includegraphics[width=1.0\columnwidth]{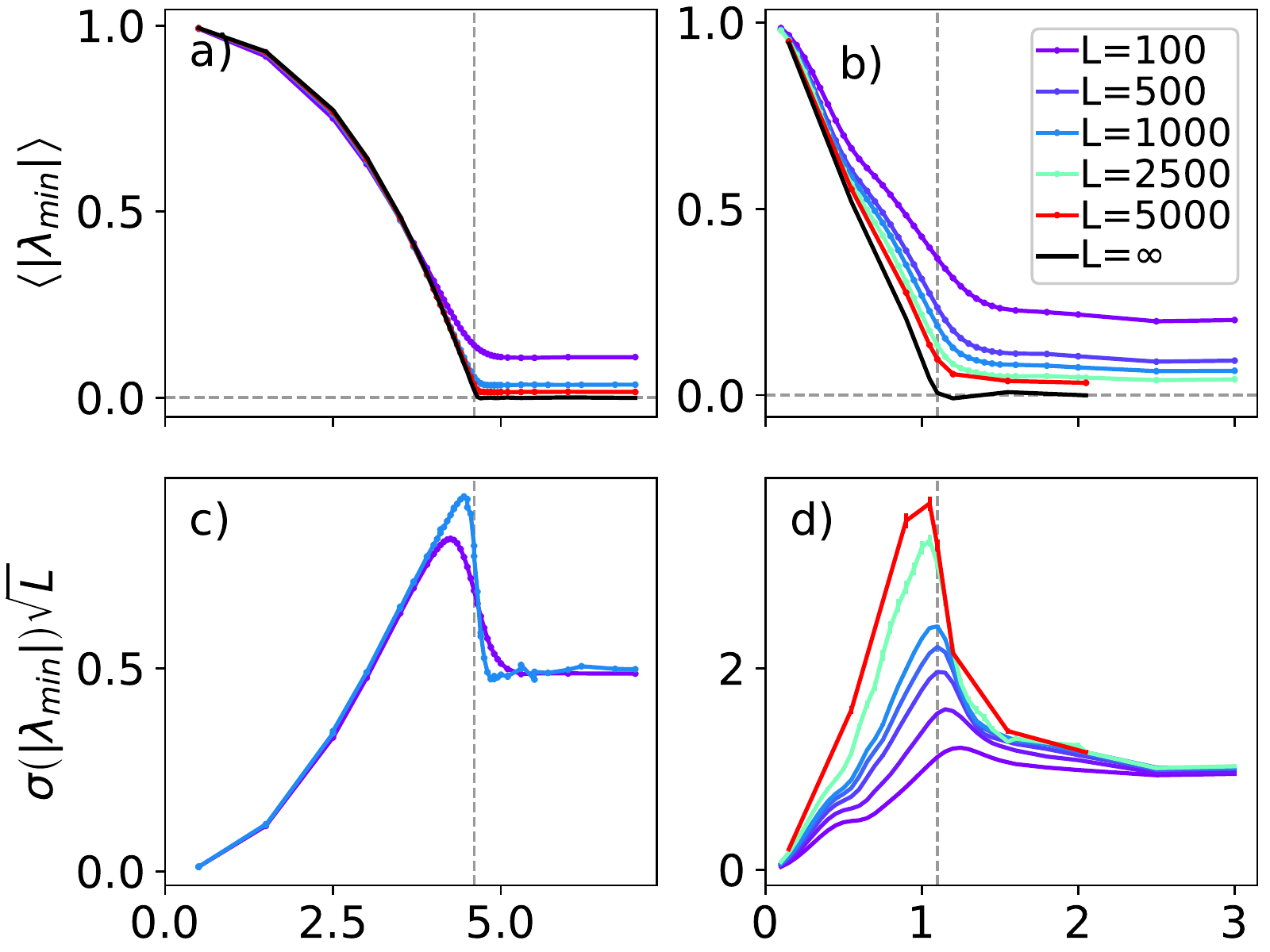}
\caption{Mean and scaled standard deviation of the distribution of
  $|\lambda_{min}|$ calculated for an ensemble of $N_s=10000$ systems
  for $L \leq 500$ and $N_s=1000$ for $L \geq 1000$. Panels a and c
  show the quench of the microscopic model also considered in
  Fig.~\ref{fig1}, while panels b and d show the quench from the
  \emph{ground state} of Hamiltonian
  \ref{microscopic_model_hamiltonian} at $h=0,J=1$ propagated with the
  same Hamiltonian at parameters $h=5,J=1$. Black lines are calculated
  by a linear extrapolation in $1/\sqrt{L}$. The vertical dashed lines
  show the estimated critical time.}
\label{mean_std_figure}
\end{figure}

The distribution of $\lambda_{min}$ is plotted in
Fig.~\ref{h_0_to_5_quench_fig}d for the $L=500$ system around the
critical time determined from the Loschmidt zeros. Indeed, we see the
smallest eigenvalues reaching zero at $t \approx 1.1$ and the
distribution undergoing a qualitative transition to a form that
closely follows the Rayleigh distribution above critical times. That
this is indeed a phase transition is demonstrated in
Fig.~\ref{mean_std_figure} for two different quenches. In
Fig~\ref{mean_std_figure}a and Fig~\ref{mean_std_figure}c we plot
$\left \langle |\lambda_{min}| \right \rangle$ and
$\sigma(|\lambda_{min}|)\sqrt{L}$ for the quench of the miscroscopic
model also considered in Fig.~\ref{fig1}. We observe a plateau in
$\left \langle |\lambda_{min}| \right \rangle$ for $t \gtrsim t_c
\approx 4.6$ consistently with the appearance of the Loschmidt zeros
in Fig.~\ref{fig1}. We have verified that $\left \langle
|\lambda_{min}| \right \rangle$ scales as $1/\sqrt{L}$ in the
large-$t$ region. A phase transition is also clearly signalled by the
jump in $\sigma(|\lambda_{min}|)\sqrt{L}$ observed at $t=t_c$. The
apparent convergence of the curve for
$\sigma(|\lambda_{min}|)\sqrt{L}$ indicates that the standard
deviation scales as $1/\sqrt{L}$ for all times. This confirms that the
eigenvalue distribution of $M(t)$ converges to a region with a
well-defined boundary.

Fig.~\ref{mean_std_figure}b and Fig.~\ref{mean_std_figure}d show the
corresponding data for a quench starting form the ground state of
model \ref{microscopic_model_hamiltonian} with $h=0,J=1$ and quenching
to $h=5,J=1$. The behaviour of $\left \langle |\lambda_{min}| \right
\rangle$ is similar with an initial decrease and a transition to a
plateau where $\left \langle |\lambda_{min}| \right \rangle$
approaches zero with increasing $L$. The phase transition point can be
estimated from the extrapolated data as $t_c \approx 1.1$. The main
difference to the previous quench is that $\sigma(|\lambda_{min}|)$
decreases slower than $1/\sqrt{L}$ for $t<t_c$, which is consistent
with the qualitatively ``fuzzier'' boundary of the eigenvalue
distribution compared to the quenches of Fig.~\ref{fig1}.
Nevertheless, for $t \gtrsim 1.1$ both the mean and the standard
deviation scale to zero, indicating that $|\lambda_{min}|=0$ for all
realizations in the thermodynamic limit.

\begin{figure}
\includegraphics[width=1.0\columnwidth]{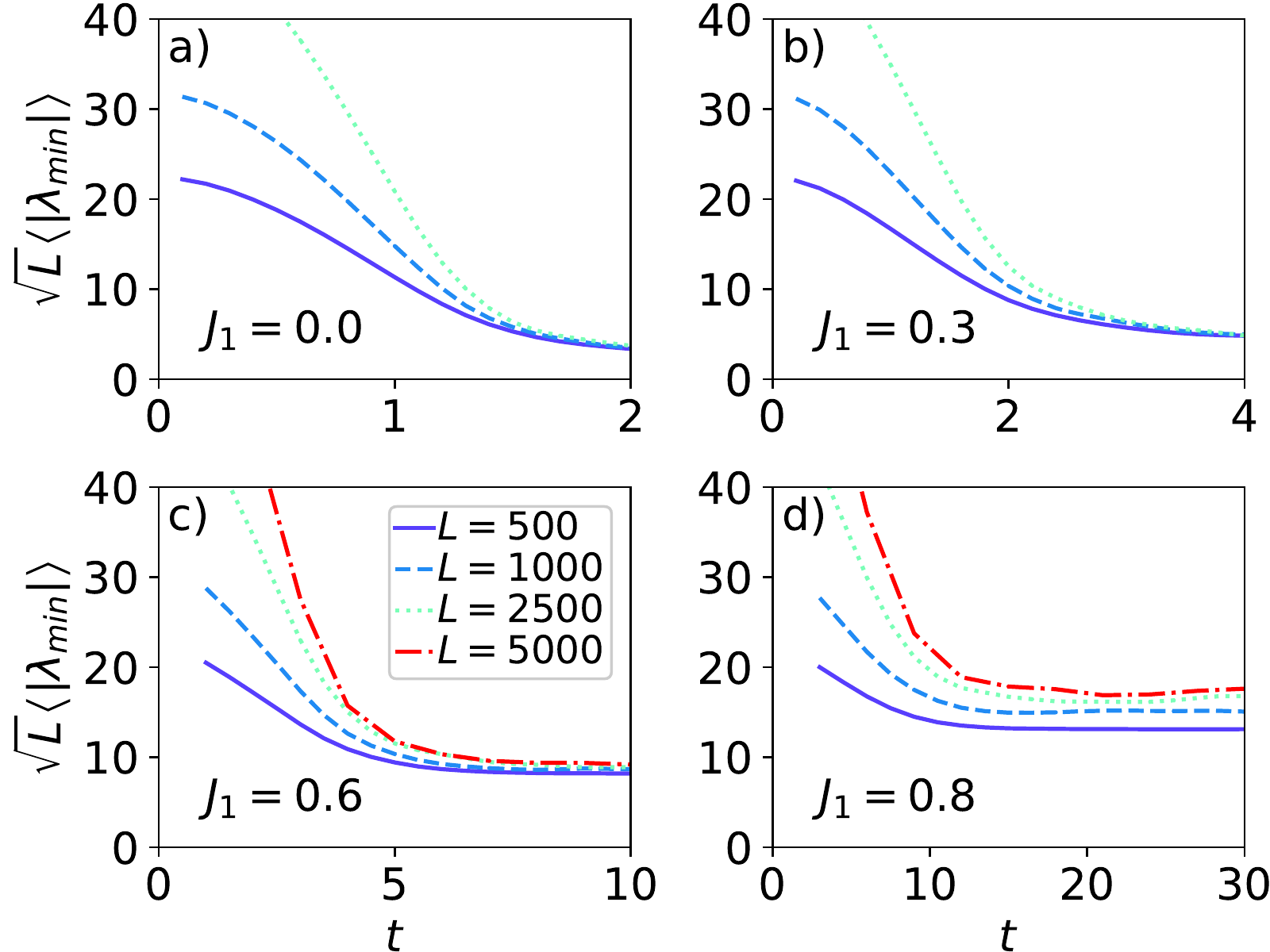}
\caption{Scaled mean of the distribution of $|\lambda_{min}|$ for
  quenches from the ground state of Hamiltonian
  \ref{microscopic_model_hamiltonian}. The system is quenched from
  $J=J_0=1,h=5$ to $J=J_1,h=5$ with the same random potential
  realization in the initial and final states. The panels show results
  for different $J_1$. Ensemble sizes are as in
  Fig. \ref{mean_std_figure}.}
\label{different_J_final_figure}
\end{figure}

\emph{Strong and weak quenches--} It has been widely observed that
sudden quench through an equilibrium critical point in the parameter
space typically results rich dynamics \cite{PhysRevX.11.031062} and
dynamical phase transitions compared to quenches confined to a same
equilibrium phase. This property, though not without exceptions
\cite{Vajna2014}, is so generic that it has been proposed even as a
diagnostic tool to investigate equilibrium phase boundaries.  A
remarkable feature of the transitions studied in this work is that
they are not related to the underlying Anderson localization
transitions. The model \eqref{microscopic_model_hamiltonian} exhibits
Anderson localization for any disorder strength $h>0$, and one might
wonder if the DQPT is related to quenching from $h=0$ to $h>0$ or vice
versa. This is not the case, as demonstrated in
Fig. \ref{different_J_final_figure}. Here the quench is between two
points in the parameter space where the system is deep in the
localized phase. However, it is evident that the scaling of $\left
\langle |\lambda_{min}| \right \rangle$ with $\sqrt{L}$ is different
for early and late times, signifying an occurrence of a DQPT. We note
that DQPTs appear also in the effective GUE model discussed above,
which does not exhibit a localization transition.

We finally consider the difference between ``weak'' and ``strong''
quenches. If we let $J_1$ approach $J_0$, the post-quench Hamiltonian
approaches the initial Hamiltonian. If the system had a gapped ground
state, we would expect that the ground states of $H_1$ and $H_0$ also
approach each other, and that any possible dynamical phase transition
would eventually disappear. However, since there is no gap in the
thermodynamic limit, this argument is not applicable. In fact, it
seems that the DQPT appears also for small quenches of $J$ but
reaching the scaling limit requires larger system sizes, as seen in
Fig. \ref{microscopic_model_hamiltonian}c and
\ref{microscopic_model_hamiltonian}d.

\emph{ Discussion and Summary---} In this work we developed the theory
of Loschmidt echo in strongly disordered non-interacting fermionic
models after a quench.  We showed that the zeros of the Loschmidt echo
are best understood in terms of the eigenvalues of the Loschmidt
matrix $M(t)$ defined in Eq. \eqref{Z_M_first_definition}, and
developed a scaling theory that links the discovered new type of
singular many-body dynamics to phase transitions in the eigenvalue
distribution of $M(t)$. Unexpectedly, we find that generic quenches
lead to qualitatively similar DQPTs. Specifically, DQPTs appear also
when performing quenches deep within the localized phase, which
clearly rules out the idea that DQPTs could be employed to pinpoint
equilibrium transitions in disordered systems. A similar DQPT was also
found in a generic GUE random matrix model, which points to a rather
universal phenomenon independent of details such as dimension as long
as disorder is present in the quench power spectrum.

Our findings constitute a radical departure from the previous results
in models exhibiting localization transitions. The bosonic Anderson
model \cite{PhysRevA.97.033624} and the fermionic Aubry-Andre model
\cite{PhysRevB.95.184201} were found to exhibit periodic Loschmidt
zeros when quenched through the localization transition point, but not
when quenched within the localized or delocalized phase. However, the
crucial difference is that Ref.~\cite{PhysRevB.95.184201} considers a
single fermion and Ref.~\cite{PhysRevA.97.033624} considers many
bosons in the same single-particle state, while in our work we deal
with generic fermionic Slater determinant states. Periodic dynamical
phase transitions were also found in the many-body time evolution of
the interacting Aubry-Andre model with system sizes up to $L=100$
\cite{PhysRevB.103.224310}. It would be interesting to revisit the
Aubry-Andre model using the eigenvalues of the $M(t)$-matrix and
perform a scaling analysis to determine the type of the zero-manifolds
as we have done here for the model with Anderson disorder.

Finally, from a methodological point of view, we expect the ideas
developed in our work to be useful for treating many non-interacting
fermionic models without translation invariance. Considering the
eigenvalues of $M(t)$ is complementary to the recently developed
cumulant method \cite{PhysRevX.11.041018,PhysRevResearch.4.033032}
which has been successfully applied to various strongly-correlated
systems in 1d and 2d. In the present work we developed a variant of
the cumulant method to efficiently study large non-interacting but
disordered models. We expect these ideas to stimulate further studies
of the dynamics of disordered systems.

\emph{Acknowledgements-- } The authors acknowledge the Academy of Finland project 331094 for support.

\bibliography{references}

\appendix
\section{Expected distribution of minimal eigenvalue}

To model the distribution of the minimal eigenvalue of the matrix
$M(t)$ for $t>t_c$, we consider a small circle of radius $R$ around
the origin and assume that the eigenvalues within this circle are
independently and uniformly distributed with density
$\rho_{\lambda}$. The number of eigenvalues within the circle is thus
$N=\pi R^2 \rho_{\lambda}$. The probability that all points are
outside a smaller circle of radius $r$ is
\begin{equation}
  P(|\lambda_{min}|>r)=(1-r^2/R^2)^N.
\end{equation}
The cumulative distribution function for the scaled variable
$|\lambda_{min}|\sqrt{\rho_{\lambda}}$ is then
\begin{equation}
\begin{split}
  F(x)&=P(|\lambda_{min}|\sqrt{\rho_{\lambda}}<x)=\\
      &=P(|\lambda_{min}|<x/\sqrt{\rho_{\lambda}})=1-P(|\lambda_{min}|>x/\sqrt{\rho_{\lambda}})\\
      &=1-\left(1-\frac{\pi x^2}{\pi \rho_{\lambda} R^2}\right)^{N}=1-\left(1-\frac{\pi x^2}{N}\right)^{N}.
\end{split}
\end{equation}
When the system size grows, we expect $\rho_{\lambda} \rightarrow
\infty$ so that $N \rightarrow \infty$. The cumulative function then
converges to
\begin{equation}
  F(x) \rightarrow 1-\exp(-\pi x^2).
\end{equation}
The cumulative function of the Rayleigh distribution is usually
written as
\begin{equation}
F_{Rayleigh}(x)=1-\exp(-x^2/(2 \sigma^2)),
\end{equation}
where $\sigma$ is a scale parameter. Thus
$|\lambda_{min}|\sqrt{\rho_{\lambda}}$ becomes Rayleigh distributed
with scale parameter $\sigma=1/\sqrt{2 \pi}$, while the distribution
of $|\lambda_{min}|$ becomes increasingly narrow as $\rho_{\lambda}$
increases, and can be approximated by a Rayleigh distribution with the
scale parameter $\sigma=1/\sqrt{2 \pi \rho_{\lambda}}$.

\section{Specialization of the cumulant method for non-interacting systems}

The cumulant method developed in \cite{PhysRevX.11.041018} can be used
to locate the Loschmidt zeros in the complex plane. In principle the
method can be directly applied also to the non-interacting models
discussed in this work. However, specializing the computation of the
cumulants to the case of Slater determinant states naturally offers a
huge numerical advantage, because it avoids handling general many-body
state vectors.

The Loschmidt echo as a function of the imaginary time $\tau$ can be
defined as
\begin{equation}
Z(\tau)=\braket{\Psi | \exp(-\tau H_1) | \Psi},
\end{equation}
where $H_1$ is the post-quench Hamiltonian and $\ket{\Psi}$ is taken
to be some eigenstate of the pre-quench Hamiltonian $H_0$. If $H_0$ is
a non-interacting fermionic Hamiltonian, then the state $\ket{\Psi}$
is a slater determinant of single-particle states. We collect the
states to a tall matrix $V$ so that each column is a single-particle
eigenstate. If we also assume that $H_1$ is non-interacting, the state
$\ket{\Psi(\tau)}=\exp(-\tau H_1) \ket{\Psi}$ is also always a slater
determinant where the single-particle states are time-developed by
$H_1$. Thus $\ket{\Psi(\tau)}$ is represented by the time-developed
tall matrix $V(\tau)=\exp(-\tau H_1) V$.

The overlap of slater determinants is given by the determinant of the
overlap matrix \cite{plasser2016efficient},
\begin{equation}
Z(\tau)=\det(V^\dagger V(\tau)).
\end{equation}
In the interacting case \cite{PhysRevX.11.041018} the cumulants are
calculated from the moments
\begin{equation}
  \mu_n=\partial^n_{\tau} Z(\tau).
\end{equation}
However, the first derivative of the determinant of a matrix
$M(\tau)=V^\dagger V(\tau)$ is
\begin{equation}
  \partial_{\tau} \det(M(\tau)) = \det(M(\tau)) \tr \left( M(\tau)^{-1} \partial_{\tau} M(\tau) \right),
\end{equation}
with successively more complicated formulas for higher derivatives. It
is thus difficult to directly compute the moments as derivatives of
the $Z$.

For the Slater determinant case we can instead start from the formula
for the cumulants
\begin{equation}
  \kappa_n = \partial^n_\tau \log(\det(M(\tau))) |_{\tau=\tau_0},
\end{equation}
which, for $n=1$ becomes
\begin{equation}
  \kappa_1=\tr( M(\tau)^{-1} \partial_{\tau} M(\tau) )|_{\tau=\tau_0}.
\end{equation}
Let us then define a matrix function $K'(\tau)$ such that

\begin{equation}
  K'(\tau) = M(\tau)^{-1} \partial_{\tau} M(\tau),
\end{equation}
or equivalently
\begin{equation}
  \partial_{\tau} M(\tau)=M(\tau) K'(\tau).
  \label{K_definition_equation}
\end{equation}
We then have that $\kappa_n=\tr(\partial_{\tau}^{n-1} K'(\tau) )|_{\tau=\tau_0}$.

In the case of a single particle $M$ would be a $1 \times 1$ matrix
and we could just define $K(\tau)=\log(M(\tau))=\log(Z(\tau))$, and
$K'(\tau)=\partial_{\tau} K(\tau)$. $K$ and $M$ would then just be the
cumulant and moment generating functions. However, we don't want to do
so because the derivative of the matrix logarithm is again
non-trivial, and we don't actually need to define $K(\tau)$. It's
enough to have a well-defined $K'(\tau)$ which is not necessarily a
derivative of a known function $K(\tau)$.

Now we can proceed as in the usual case to derive a formula that
relates cumulants to moments. But now our moments and cumulants are
matrices,
\begin{equation}
  K'(\tau)=\sum_{n=1}^{\infty} K_{n} \frac{(\tau-\tau_0)^{n-1}}{(n-1)!},
\end{equation}
and
\begin{equation}
  M(\tau)=\sum_{n=0}^{\infty} M_n \frac{(\tau-\tau_0)^n}{n!}.
\end{equation}
Equating the coefficient of $\tau^{n-1}$ on both sides of equation
\ref{K_definition_equation} we get
\begin{widetext}
\begin{equation}
  \frac{M_n}{(n-1)!}=\sum_{k=1}^n M_{n-k} K_{k} \frac{1}{(k-1)! (n-k)!} = \frac{M_0 K_n}{(n-1)!} + \sum_{k=1}^{n-1} M_{n-k} K_{k} \frac{1}{(k-1)! (n-k)!}
\end{equation}
\end{widetext}
which can be solved for $K_n$ as
\begin{widetext}
\begin{equation}
  K_n=M_0^{-1} \left( M_n - \sum_{k=1}^{n-1} M_{n-k} K_{k} \frac{(n-1)!}{(k-1)! (n-k)!} \right) =
  M_0^{-1} \left( M_n - \sum_{k=1}^{n-1} \binom{n-1}{k-1} M_{n-k} K_{k} \right).
\end{equation}
\end{widetext}
For the $1 \times 1$ case this again just reduces to the usual
recursive formula for the cumulants in terms of the moments. In fact,
it is the same formula, but it has to be remembered that the matrices
do not necessarily commute.

Numerically, we first compute the moment matrices $M_n=V^{\dagger}
\partial_{\tau}^n V(\tau) |_{\tau=\tau_0}$, and then use the recursive
formula to get the cumulant matrices $K_n$. The cumulants $\kappa_n$
are then found as the trace of the matrices $K_n$. Finding the zeros
using the cumulants proceeds as in the interacting case
\cite{PhysRevX.11.041018}.

\label{supplementary_material}

\end{document}